\documentclass [12pt]{article}
\usepackage {amssymb}
\usepackage {amsmath}
\usepackage {graphicx}

\begin{document}
{ \footnotesize{\noindent in:  \textit{Einstein and Poincar{\'e}:
The Physical Vacuum}, Ed.: V. V. Dvoeglazov (Apeiron, Montreal,
2006), pp. 143-153. }}

\section*{The Tessellattice of Mother-space as
a Source and Generator of Matter and Physical Laws}

\bigskip
\begin{center}
\textbf{Volodymyr Krasnoholovets}
\end{center}

\bigskip

\noindent Institute for Basic Research, 90 East Winds Court, Palm Harbor, FL
34683, U.S.A. \ \ \textit{E-mail address}: \  v${\! \_}{\kern
1pt}$kras@yahoo.com

\quad\quad\quad\quad\quad\quad\quad\quad\quad\quad\quad\quad\quad
\quad\quad\quad\quad\quad\quad\quad\quad\quad\quad\quad\quad {1
December 2004}

\bigskip
\bigskip
{\small \textbf{Abstract.} Real physical space is derived from a mathematical space constructed as a tessellation lattice of primary balls, or superparticles. Mathematical characteristics, such as distance, surface and volume, generate in the fractal tessellation lattice the basic physical notions (mass, particle, the particle's de Broglie wavelength and so on) and the corresponding fundamental physical laws. Submicroscopic mechanics developed in the tessellattice easily results in the conventional quantum mechanical formalism at a larger atomic scale and Newton's gravitational law at a macroscopic scale.}

\bigskip
\bigskip

\subsection*{1. Introduction}

\hspace*{\parindent} Although Poincar\'e (1905a) was the first to write the relativistic transformation law for charge density and velocity of motion, Einstein's (1905) special relativity article received wide recognition perhaps due to his introduction of a radically new abstract approach to fundamentals, which then culminated in his famous theory of general relativity (Einstein, 1916). Due to his great predictions, which were verified experimentally, abstract theoretical concepts took root in the minds of a majority of physicists. Einstein's approach resembled rather a generalized description that descended to particulars through a series of postulated axioms. His general relativity considers how matter and geometry, constructed in the empty space, coexist and influence each other, though matter is not an intrinsic property of space.

Einstein's thoughts regarding an aether were expressed in his well-known lecture (Einstein, 1920). He noted since space was endowed with physical qualities, an aether existed. Then he mentioned, according to general relativity, space without an aether is unthinkable (light would not propagate, there would not any space-time intervals in the physical sense, etc.). Nevertheless, Einstein stressed that this aether might not be thought of as endowed with quality characteristic of a ponderable medium, consisting of parts that might be tracked through time. However, the basic issue remained: Why could the aether not be associated with a substrate? This was never clarified by Einstein completely.

At the same time, a hypothesis about the existence of an aether as a material substrate responsible for electromagnetic wave propagation has been tested by many researchers (Miller, 1933, Essen, 1955, Azjukowski, 1993). A new optical method of the first order was proposed and implemented by Galaev (2002) for measurements of the aether-drift velocity and kinematic viscosity of aether. Galaev's results correlate well with the results of other researchers quoted above.
Observability, reproducibility and repeatability of aether drift effects have been conducted in various geographical conditions with help of different methods of measurements and in various ranges of electromagnetic waves. Overall, the above-mentioned researches strongly supported the idea the aether is a subsrate responsible for propagation of electromagnetic waves. These studies shed light on negative results of measurements of aethereal wind by Michelson and Morley: Their tool had too low a sensitivity.

Other researchers have made a demonstration of direct interaction of matter with a subquantum medium. In particular, an influence of a new ``strange" physical field on specimens was fixed by Baurov (2002), Benford (2002) and Urutskoev et al. (2002). Similar effects are described by Shipov (1997), though the changes in samples examined were associated with the so-called ``torsion radiation" introduced by Shipov as a primary field that allegedly was dominating over a vague physical vacuum long before its creation. One more incomprehensible phenomenon is the Kozyrev effect (Kozyrev and Nasonov, 1978) at which a bolometer centrally located in the focal point of a telescope recodes a signal from a star much earlier than the light signal hitting the focal point.

Let us look now briefly at Poincar\'e's studies. His researches were also highly abstract especially those dealing with mathematics and mathematical physics. Nevertheless, in physics applications, he tried to bear on natural laws as close as possible. Granted, Poincar\'e (1905b, 1906) believed any new success in science was an additional support of determinism. In his works, he tried to start just from details that then should disclose the problem studied as a whole. Poincar\'e (1905b) strongly supported the idea of an aether, considering motion of a particle as accompanied by an aether perturbation. Such idea, perturbation of the aether by a moving object, dominated over leading mathematicians and physicists up to beginning of the 20th centuriy.

Therefore, his idea deserves credit (if a kind of an aether in fact exists). Poincar\'e considered particles as peculiar points in the aether, though he did not develop further ideas on its construction nor principles of the motion of material objects in it. Experimental facts were not abundant at that time as well as theoretical elaborations of condensed matter physics, which would help one to look at a possible theory of aether in more detail. Besides, mathematical methods of description of space were also rather in an embryonic state at beginning of the 20th century despite the fact it was Poincar\'e who proposed and developed new concepts and methods of the investigation of space as such. At that time, facts were not numerous as now, which then did not allow Poincar\'e to consolidate ideas on space and aether in a unified generalized concept of real space.

However, today when practically all the facts are already available, we may try to look at a possibility of unification of mathematical and physical ideas regarding incorporation of an aether with space in one unified object of comprehensive study.

\subsection*{2. The constitution of space}

\hspace*{\parindent} Many researchers are involved in the search for a theory of everything (TOE). However, how about a ``theory of something"? The problem was studied by Bounias (2000) on the basis of pure mathematical principles. He firmly believed the ultimate theory could be some mathematical principle.

Following upon Bounias (2000), Bounias and Krasnoholovets (2002, 2003),
we can explore the problem of the constitution of space in terms
of topology, set theory and fractal geometry. Evidently according
to set theory, only the empty set (noted $\O$) can represent
nearly nothing. If \textit{F} is a part of \textit{E}, then the
remaining part of \textit{E} that does not contain \textit{F} is
the complementary of \textit{F} in \textit{E}, which is noted as
$\complement _{{\kern 1pt} {\kern 1pt} E} \left( {F} \right).$ The
empty set $\O $ is contained in any set \textit{E}, i.e.
$\complement _{{\kern 1pt} {\kern 1pt} E} \left( {E} \right) = \{
\O ,{\kern 1pt} {\kern 1pt} {\kern 1pt} {\kern 1pt} {\kern 1pt}
E\} $, then $\complement _{{\kern 1pt} E} \left( {E} \right) = \O
$; this last result together with $\complement _{{\kern 1pt} E}
\left( {\O} \right) = E$ are known as the first law of Morgan. All
this allowed Bounias to conclude the complement of the empty
set is the empty set: $\complement _{{\kern 1pt} {\kern 1pt} \O}
\left( {\O} \right) = \O $. Following von Neumann, Bounias
considered an ordered set, \{\{\{$\O $,\{$\O $\}\}, \{$\O $,\{$\O
$,\{$\O $\}\}, and so on. Looking at the set, one can count its
members: $\O $= zero, \{$\O $\}, \{$\O $\}\} = one, and so on.
This is the empty set as long as it consists of empty members and
parts. However, on the other hand, it has the same numbers of
members as the set of natural integers, $N = \{
0,\,\,\,1,\,\,\,2,\,\,...,\,\,n\} $. Although it is properly that
the reality is not reduced to the enumeration, the empty sets give
rise to mathematical space that in turns brings about physical
spaces. So {\it something} can exist emerging from emptiness.

The empty set is contained in itself, hence it is a
non-well-founded set, or hyperset, or empty hyperset. Any parts of
the empty hyperset are identical, either a big part ($\O$) or the
singleton $\{ \O \} $; the reunion of empty sets is also the same:
$\O \,\,\, \cup \,\,\,\left( {\O } \right)\,\,\, \cup \,\,\{ \O \}
\,\,\, \cup \,\,\{ \O ,\,\,\{ \O \} \} \,\, \cup ...{\kern 1pt}
{\kern 1pt} {\kern 1pt} {\kern 1pt} \, = \,\,\,\O $. This is the
major characteristic of a fractal structure, which means the
self-similarity at all scales (in physical terms, from the
elementary sub atomic level to cosmic sizes). One empty set $\O $
can be subdivided into to two others; two empty sets generate
something $\left( {\O} \right) \cup \left( {\O} \right)$ that is
larger than the initial element. Consequently, the coefficient of
similarity is $\rho \in \left[ {{\raise0.5ex\hbox{$\scriptstyle
{1}$}\kern-0.1em/\kern-0.15em\lower0.25ex\hbox{$\scriptstyle
{2}$}},\,\,1} \right]$. In other words, $\rho $ realizes the
fragmentation when it falls within the interval
$]{\raise0.5ex\hbox{$\scriptstyle
{1}$}\kern-0.1em/\kern-0.15em\lower0.25ex\hbox{$\scriptstyle
{2}$}},\,\,1 [ $   and the union when $\rho$ with the interval
$]{\kern 1pt} 0,{\kern 1pt} {\kern 1pt} {\kern 1pt} {\raise0.5ex
\hbox{$\scriptstyle{1}$}\kern-0.1em/\kern-0.15em\lower0.25ex
\hbox{$\scriptstyle{2}$}}]$ yields $ ]{\kern 1pt} 0,{\kern 1pt}
{\kern 1pt} {\kern 1pt} {\kern 1pt} 1 [$. The coefficient of
similarity allows one to estimate the fractal dimension of the
empty hyperset, which owing to the interval $]{\kern 1pt} 0,{\kern
1pt} {\kern 1pt} {\kern 1pt} {\kern 1pt} 1 [ $ becomes a fuzzy
dimension.

Time can be called ``nothing", because it is a singleton that does
not have parts (in other case it will be in contradiction to the
definition of time as such). The nothingness singleton ($ \in $) is
absolute unique. It is the greatest lower boundary of everything
that exists; this is the infimum of existence, Figure 1.

\begin{figure}
\begin{center}
\includegraphics[scale=2.9]{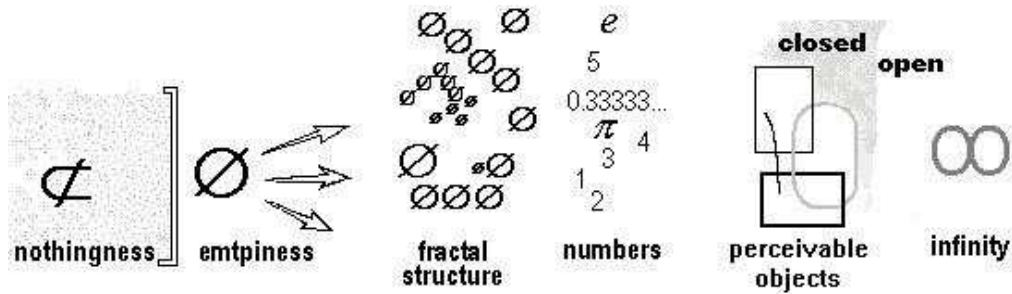}
\caption{\small  Range of things, from non-existence to something
empty whose structure gives rise to something non-empty and up to
infinity (from Bounias and Krasnoholovets, 2002).}
   \label{Figure 1}
\end{center}
\end{figure}

4-D mathematical spaces have parts in common with 3-D spaces, which
gives 3-D closed structures; then there are parts with 2-D, 1-D and
dimension zero (points). General topology indicates the origin of
time, which should be treated as an assembly of sections $S_{{\kern
1pt} i} $ of open sets. Indeed, fractality of space generates fuzzy
dimensions (Bounias and Krasnoholovets, 2003a) and hence a part in
common of a couple of open sets $W_{m} $ and $W_{n} $ with different
dimensions\textit{ m} and \textit{n} also accumulates points of the
open space. If $m > n$, then those points that belong to $W_{m} $
and would not belong to the section of the given sets cannot be
included in a {\it x}-D object. Bounias and Krasnoholovets (2002)
exemplified this in the following way: ``You cannot put a pot into a
sheet without changing the shape of the 2-D sheet into a 3-D packet.
Only a 2-D slice of the pot could be a part of a sheet." Therefore
infinitely many slices, i.e. a new subset of sections with
dimensionality from 0 to 3, ensure the raw universe in its timeless
form.

Thus a physical space one can provide by closed intersections
(timeless Poincar\'{e} sections) of abstract mathematical spaces.
What would happen with these sections $S_{i} $ that all belong to
an embedding 4-space? A series of sections $S_{i} $, $S_{i{\kern
1pt} {\kern 1pt} + {\kern 1pt} {\kern 1pt} 1} $, $S_{i + {\kern
1pt} {\kern 1pt} 2} $, etc. resembles the successive images of a
movie and only nothing does not move. Because of that, difference
of distribution of objects within two corresponding sections will
mean a detectable increment of time. Therefore, time will emerge
from order relations holding on these sections.

Two successive slices show a characteristic of mathematical objects
from one to the next section. In other words, this is a mapping. The
first section produces some \textit{x} that then becomes $f(x)$ on
the next one. The mapping between nearest sections can be treated in
the framework of an indicatrix function $1{\kern 1pt} {\kern 1pt}
(x)$ and Uryson's theorem. By definition, $1{\kern 1pt} (x)$ for any
\textit{x} states yields $1{\kern 1pt} (x) = 0$ if \textit{x} has
one property and $1 {\kern 1pt} (x) = 1$ if \textit{x} has an
alternative property. A combination of$f(x)$ with $1\left( f(x)
\right)$ makes a demonstration that the result is depend on whether
the variable \textit{x} belonging to one part of the frame in $S_{i}
$ or will belong to the same part in $S_{{\kern 1pt} i + {\kern 1pt}
1} $. The complete function is a composition of the variables with
their distribution. In other words, the function has the structure
of a moment; it was called a `moment of junction' \textit{MJ} by
Bounias (2000), Bounias and Krasnoholovets (2003b). The function
\textit{MJ} describes the smallest increment of space (only one
point is not at the same topological position for \textit{MJ} to
permit the change). Furthermore, such a fine change of \textit{MJ}
defines also an increment in time - the minimal change. Since there
is no any thickness between two sections $S_{{\kern 1pt} i} $ and
$S_{{\kern 1pt} i{\kern 1pt} {\kern 1pt} + {\kern 1pt} {\kern 1pt}
1} $, the moment of junction \textit{MJ} rigorously describes a
differential element of space, which is also a differential element
of time. And this validates differential geometry for the
description of the Universe.

\subsection*{3. Measure, distance, metric and objects}

\hspace*{\parindent} The concept of measure usually involves such
particular features as the existence of mappings and the indexation
of collections of subsets on natural integers. Classically, a
measure is a comparison of the measured object with some unit taken
as a standard. The ``unit used as a standard", this is the part
played by a gauge (\textit{J}).

A measure involves respective mappings on spaces that must be
provided with the rules $\cap , {\kern 1pt} {\kern 1pt} \cup $ and
$\complement $. According to Bonaly and Bounias (1996), in spaces
of the $\mathbb{R}^{{\kern 1pt} n}$ type, tessellation by balls
are involved which again demands a distance to be available for
the measure of diameters of intervals. The intervals can be
replaced by topological balls and therefore the evaluation of
their diameter still needs an appropriate general definition of a
distance. More comprehensive determinations of measure, distance,
metric and objects, which involve topology, set theory and fractal
geometry have been done by Bounias and Krasnoholovets (2003a,
2004).

In physics, a ruler is called a metric. As a rule, in mathematics spaces
including topological spaces were treated as not endowed with a metric and
the properties of metric spaces have not been the same as those of
non-metric spaces. However, in 1994 Bounias (see e.g. Bonaly and Bounias,
(1996)) could show that there was not exist a non-metric topological space!

Indeed, union and intersection allow the introduction of the symmetric
difference between two sets $A_{{\kern 1pt} i} $ and $A_{{\kern 1pt} j} $
\begin{equation}
\label{eq1} \Delta \left( {A_{{\kern 1pt} i},{\kern 1pt}
A_{{\kern 0.3pt} j}}  \right) = \mathop {\complement} \limits_{
\cup {\kern 1pt} {\kern 1pt} \{ A_{{\kern 1pt} i} \}} \cup
_{{\kern 2pt} i \ne j} \left( {A_{{\kern 1pt} {\kern 1pt} i} \cap
A_{j}}  \right)
\end{equation}

\noindent i.e. the complementary of the intersection of these sets
in their union. The symmetric difference satisfies the following
properties: $\Delta \left( {A_{{\kern 1pt} i} ,{\kern 1pt} {\kern
1pt} {\kern 1pt} A_{j}}  \right) = 0$ if $A_{{\kern 1pt} i} =
A_{j} $, $\Delta \left( {A_{{\kern 1pt} i}, {\kern 2pt} A_{j}}
\right) = \Delta \left( {A_{j} ,{\kern 1pt} {\kern 1pt} {\kern
1pt} A_{{\kern 1pt} i}}  \right)$ and $\Delta \left( {A_{{\kern
1pt} i} ,\,\,A_{j}}  \right)$ is contained in the union of $\Delta
\left( {A_{{\kern 1pt} i} ,\,\,A_{j}} \right)$ and $\Delta \left(
{A_{{\kern 1pt} j} ,\,\,A_{k}} \right)$. This means that it is a
true distance and it can also be extended to the distance of
three, four and etc. sets in one, namely, $\Delta \left(
{A_{{\kern 1pt} i} ,\,\,A_{j} ,\,\,A_{{\kern 1pt} k} ,{\kern 1pt}
{\kern 1pt} {\kern 1pt} \,{\kern 1pt} A_{{\kern 1pt} l} ,{\kern
1pt} {\kern 1pt} {\kern 1pt} \,{\kern 1pt} ...} \right)$. Since
the definition of a topology implies the definition such a set
distance, every topological space is endowed with this set metric.
The norm of the set metric is $||A||\,\, = \,\Delta \left( {\O ,A}
\right)$. Therefore, all topological spaces are metric spaces,
$\Delta $-metric spaces, and they are measurable.

Let us look now at the remaining part, i.e. the intersection of
the sets. If they are of unequal dimensions, this intersection
will be closed, i.e. the intersection in a closed space is closed,
$ \cup _{{\kern 1pt} i \ne j} \left( {A_{{\kern 1pt} i{\kern 1pt}}
\cap A_{j}}  \right)$, which signifies the availability of
physical objects. As distances $\Delta $ are the complementariness
of objects, the system stands as a manifold of open and closed
subparts. This procedure subdivides the Universe into two parts:
the distances and the objects.

In general, we can imagine the universe as an immense drop
containing $\mathfrak{N}$ balls. Since the measure embraces such
notions as length, surface and volume, we may represent $\ell $ --
the loop distance of the universe (i.e. the perimeter that can be
measured with a ruler) -- through the parameters of those
$\mathfrak{N}$ balls. Indeed, let $\mathfrak{m}$ be the measure of
the balls (length, surface, or volume of dimension $\delta $
depending on what kind of the characteristics we are interested
in). Inside of a universe of dimension \textit{D} we have
$\mathfrak{N}$ times $\mathfrak{m}^{\delta} $ approximately equal
to $\ell {\kern 1pt} ^{D}$, so that
\begin{equation}
\label{eq2} D \sim \left( \delta  {\kern 1pt} \cdot \log
\mathfrak{m} + \log\mathfrak{N} \right) {\kern 1pt} / {\kern 1pt}
 \log \ell .
\end{equation}

Thus if we know component parts of the universe, i.e. can describe
sizes and shapes of the topological balls, we will be able to
reconstruct the large unknown structure.

\subsection*{4. Tessellation lattice of primary balls}

\hspace*{\parindent} Let us now examine what is space-time in the
approach proposed by Bounias and Krasnoholovets (2003, 2004). What
he proposed initially was the founding element. Namely, it is
generally recognized that in mathematics some set does exist. A
weaker form can be reduced to the existence of the empty set. If one
provides the empty set ($\O $) with the combination rules ($ \in
,{\kern 1pt} \subset $) and the property of complementary
($\complement $), a magma can be defined. Those preliminaries
allowed Bounias to fortunate the following theorem. The magma $\O
^{{\kern 1pt} \O} = \{ \O ,{\kern 1pt} {\kern 1pt} {\kern 1pt}
{\kern 1pt} \complement \} $ constructed with the empty hyperset and
the axiom of availability is a fractal lattice. Writing ($\O $)
denotes that the magma reflects the set of all self-mappings of $\O
$. The space constructed with the empty set cells of the magma $\O
^{{\kern 1pt} \O} $ is a Boolean lattice and this lattice
$\mathfrak{S}\left( {\O}  \right)$ is provided with a topology of
discrete space, Bounias and Krasnoholovets (2003a). A lattice of
tessellation balls has been called a ``tessel-lattice" and hence the
magma of empty hyperset becomes a fractal tessel-lattice.

The introduced lattice of empty sets ensures the existence of a
physical-like space. In fact, the consequence of spaces $\left(
{W_{m}} \right),\,\,\left( {W_{n}}  \right),\,\,...$ formed as
parts of the empty set $\O $ shows the intersections that have
non-equal dimensions, which gives rise to spaces containing all
their accumulating points forming closed sets (Bounias and
Krasnoholovets, 2003a). If morphisms are observed, then this
enables the interpretation as a motion-like phenomenon, when one
compares the state of a section with the state of mapped section.
A space-time-like sequence of Poincar\'{e} sections is a
non-linear convolution of morphisms (Bounias and Krasnoholovets,
2003a). Therefore, our space-time becomes one of the
mathematically optimum ones. And time is an emergent parameter
indexed on non-linear topological structures guaranteed by
discrete sets. In other words, the foundation of the concept of
time is the existence of order relations in the sets of functions
available in intersect sections.

Thus time is not a primary parameter. And the physical universe
has no more beginning: time is just related to ordered perceptions
of existence, not to existence itself. The topological space does
not require any fundamental difference between reversible and
steady-state phenomena, nor between reversible and irreversible
process. Rather relation orders simply hold on non-linearity
distributed topologies, and from rough to finest topologies.

Such fundamental notions as point, distance and similitude allow the
introduction of relative scales in the empty-set lattice, i.e. the
tessel-lattice, and therefore space everywhere becomes quantic
[26,28]. Indeed, from mapping $G:N^{{\kern 1pt} D} \mapsto Q$ of
$\left( {N\,\, \times \,\,N\,\, \times \,\,N\,\, \times \,\,...}
\right)$ in \textit{Q} we can identify a set of rational intervals.
In this way, for \textit{n} integers in each one of the 2-D space,
$n \times n$, the pair $\left( {1,\,\,n} \right)$ yields fraction
$1/n$ and the pair $\left( {1,{\kern 1pt} {\kern 1pt} {\kern 1pt}
{\kern 1pt} {\kern 1pt} n - 1} \right)$ yields ratio ${\kern 1pt}
1/\left( {n - 1} \right)$. So their distance is the smaller
interval, i.e. the difference between these two fractions gives the
smaller interval proportional to $1/n^{2}$ or more exactly, the
interval $1/\left( {n - 1} \right)\,\, - \,\,1/n{\kern 1pt} {\kern
1pt} {\kern 1pt} {\kern 1pt} {\kern 1pt} = {\kern 1pt} {\kern 1pt}
{\kern 1pt} {\kern 1pt} {\kern 1pt} 1/\left( {n\,\left( {n - 1}
\right)} \right){\kern 1pt} {\kern 1pt} {\kern 1pt} $ that denotes a
special scale limit depending on the size of the considered space
(recall this smaller interval, which is formed by $n^{{\kern 1pt}
2}$ grains, is constructed from $\O $). In 3-D, we will have
interval $1/n^{2}\left( {n - 1} \right)$.

Predictable orders of size from \textit{x} = 1 to \textit{x} = 60 to
be clusters/universes whose objects range from 1 (the Planck scale,
the size of an elementary cell of the tessel-lattice), to $\sim
$10$^{{\kern 1pt}10}$ elementary cells (roughly comply with
quark-like size), to about 10$^{{\kern 1pt}17}$ cells (atomic size),
to 10$^{{\kern 1pt}21}$ cells (molecular size), to 10$^{{\kern
1pt}28}$ (human size), to 10$^{{\kern 1pt}40}$ cells (star system
size) up to 10$^{{\kern 1pt}56}$ cells (greatest cosmic structures).
So, we can see that the universe suggests a quite different
organization of matter at different scales.

\subsection*{5. Generation of matter}

\hspace*{\parindent} Nowadays quantum and particle physics are
considered as most fundamental disciplines. They examine the
behavior of quantum systems, such as the interaction between
particles in the presence of this or that potential(s),
transformations of particles to the other ones, etc. However,
fundamental notions quantum physics operates with (mass, wave
$\psi$-function, wave-particle, de Broglie and Compton wavelengths,
spin and so) are out of any comprehension of their nature and
origin, because these microscopic parameters a priory are treated as
basic, or primordial. Such a viewpoint makes it possible to raise a
question about conceptual difficulties of quantum mechanics
(Krasnoholovets, 2004). So, are we able to develop deeper first
principles that will derive the fundamental notions basing on a sub
microscopic concept? And hence the ``strangeness" of quantum
mechanical behavior of particles will be complete clarified owing to
inner determinism that establishes very peculiar links in quantum
systems, which are hidden under the crude orthodox quantum
formalism. In quantum electrodynamics, neither an electric charge
nor a magnetic charge has yet been in physical terms. They are
abstract concepts transcribed into observable properties and
reflected in modelling equations.

If we wish to provide an insight of the structure of an abstract
physical vacuum, we must assume that this substance is rather
nothing, instead of being complete empty. But nothing allows the
consideration in terms of space, namely, topology, set theory and
fractal geometry, which has just been demonstrated in the previous
sections.

One of our starting points is the idea that the organization of
matter at the microscopic (atomic) level should reproduce some
submicroscopic space ordering. This means that the lattice of a
crystal should be the reflection of the arrangement of the real
physical space. This space can fully be associated with the
tessel-lattice of densely packed balls, or superparticles. And this
is the degenerate space (that one may associate with an abstract
physical vacuum). Superparticles that constitute founding cells of
the tessel-lattice are stacked without any unfilled place between
them, which refers to the nothingness singleton, addressed in the
section 2.
\begin{figure}
\begin{center}
\includegraphics[scale=2.4]{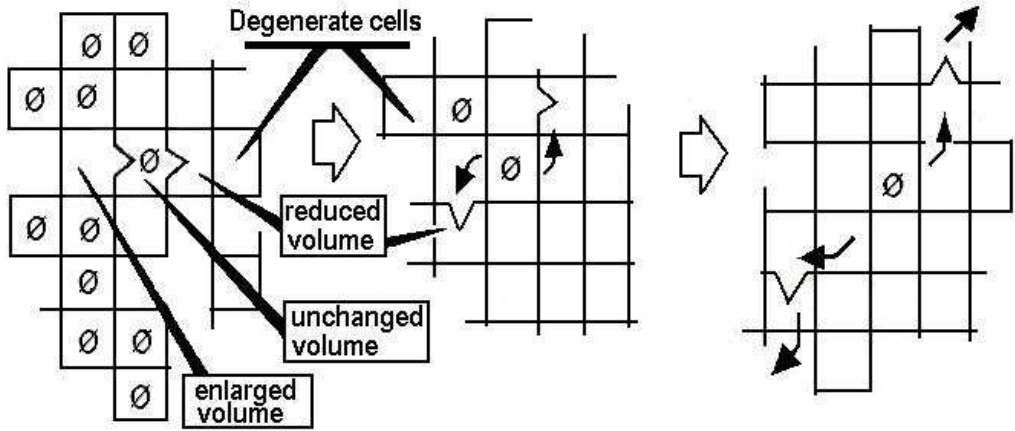}
\caption{\small  Volumetric fractality of cells as elementary
deformations of the tessellattice. These deformations can occur
with and without change in the volume of cells. Local deformations
that produce reduction of the volume of cells are associated with
the local generation of mass. These deformations can migrate in
the tessellattice from cell to cell (from Bounias and
Krasnoholovets, 2002).}
   \label{Figure 2}
\end{center}
\end{figure}

The degeneration of a cell is removed when the cell receives
several deformations, such that its volume may be reduced, while
the equivalent volume is redistributed among other cells. This
means that in terms of conventional physics the deformed
superparticle becomes a massive particle. The mass \textit{m} of
this particle is the product of a constant (\textit{C}) for
dimension assessment by ratio of the volume (\textit{V}) of a
superparticle to that of our reduced superparticle (which is
already the particle),
\begin{equation}
\label{eq3} m = C{\kern 1pt} {\kern 1pt} {\kern 1pt} V_{\rm super}
/{\kern 1pt} {\kern 1pt} {\kern 1pt} V_{\rm part} .
\end{equation}

By analogy with the deformation of a crystal lattice in the
surrounding of a foreign particle, we have to recognize that a
deformation coat appears around the particle, Figure 3. The radius
of this coat is associated with the Compton wavelength $\lambda
_{\rm {\kern 1pt} Com}$ of the particle.

So, having the particle one may try to construct its mechanics in
the tessellation space, which immediately will mean the
development of physical laws and physics in general. Since the
space should be densely packed with balls, any motion of a chosen
(deformed) ball should be expressed in terms of interaction with
other balls of the space. This brings about a radically new
approach to the behavior of matter.

\begin{figure}
\begin{center}
\includegraphics[scale=1.7]{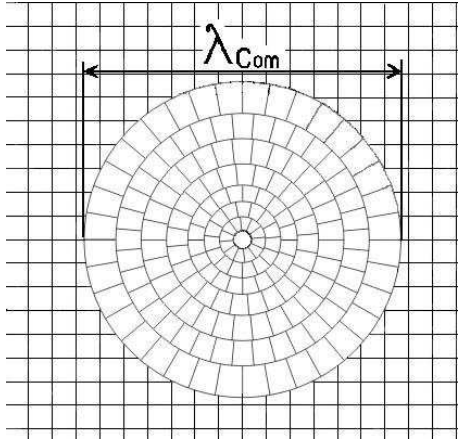}
\caption{\small  Particle as a local deformation of the
tessellattice (the central cell) and the deformation coat that
screens the particle from the degenerate tessellattice.}
   \label{Figure 3}
\end{center}
\end{figure}

\subsection*{6. The submicroscopic mechanics}

\hspace*{\parindent} The submicroscopic mechanics of particles has
been developed by the author in a series of works (see, e.g.
Krasnoholovets, 2002a). When a particle starts to move, it undergoes
``friction" on the side of the tessel-lattice and hence a packet of
deformations goes forward the particle. These elementary excitations
that migrate from cell to cell of the tessel-lattice in fact
represent a resistance of the space tessel-lattice, i.e. inertia,
and, because of that, they have been called \textit{inertons.} Thus,
collision-like phenomena are produced: deformations (inertons) go
from the particle to the surrounding space and then due to elastic
properties of the tessel-lattice some come back to the particle.
Such kind of the motion can be described by the appropriate
Lagrangian that can be written as follows (simplified here)
\begin{equation}
\label{eq4} L = \tfrac{{1}}{{2}}m{\kern 1pt} \dot {x}{\kern
1pt}^{2} + \tfrac{{1}}{{2}}\mu {\kern 1pt} \dot {\chi} {\kern 1pt}
^{2} - \sqrt {m\mu} {\kern 1pt} {\kern 1pt} {\kern 1pt} \dot
{x}\chi {\kern 1pt} {\kern 1pt} /T
\end{equation}

\noindent
where \textit{m}, \textit{x} and $\mu $, $\chi $ are the mass and the
position of the particle and its inerton cloud, respectively; 1/\textit{T}
is the frequency of collisions between the particle and the cloud.

The Euler-Lagrange equations show the periodicity in the behavior
of the particle. Namely, the particle velocity oscillates between
the initial value $\upsilon $ and zero along each section $\lambda
$ of the particle path. This spatial amplitude is determined as
follows: $\lambda = \upsilon {\kern 1pt} {\kern 1pt} T$. The same
occurs for the cloud of inertons: $\Lambda = c{\kern 1pt} {\kern
1pt} T$. So, these two amplitudes become connected by means of the
relationship $\Lambda = \lambda {\kern 1pt} {\kern 1pt} {\kern
1pt} c/\upsilon $.

Furthermore, the solutions to the equations of motion show that the
motion of particle in the tessel-lattice is characterized by the two
de Broglie's relationships for the particle: $E = h{\kern 1pt} \nu $
and $\lambda = h/\left( {m{\kern 1pt} \upsilon} \right)$ where $\nu
= 1/\left( {2T} \right)$. However, having these relationships we can
readily derive the Schr\"odinger equation. This means that at this
stage the submicroscopic mechanics passes into conventional quantum
mechanics.

The amplitude of spatial oscillations of the particle ($\lambda $)
appears in quantum mechanics as the de Broglie wavelength. The
amplitude of the particle's cloud of inertons ($\Lambda $) becomes
implicitly apparent through the availability of the wave
$\psi$-function. Therefore, the physical meaning of the
$\psi$-function becomes complete clear: it describes the range of
space around the particle perturbed by its inertons.

The next stage is that inertons transfer not only inertial, or
quantum mechanical properties of particles, but also gravitational
properties, because they transfer fragments of the deformation of
space (i.e. mass) induced by the particle. The corresponding study
(Krasnoholovets, 2002b) shows that availability of dynamic
inertons allows the derivation of Newton's static gravitational
law, 1/\textit{r}. This physical law emerges owing the fact that
the behavior of the object's inertons obeys the spreading of a
standing spherical wave that is specified by the dependency
1/\textit{r}.

\subsection*{7. Concluding remarks}

\hspace*{\parindent}Thus mysteries of quantum mechanics might
have met here a description in the real space and the inertons
have been experimentally detected in conditions predicted by the theory
(see, e.g. Krasnoholovets, 2002a). The submicroscopic mechanics fully
restores determinism. In addition, quite recently my colleagues and
I have launched the project entitled ``The inerton astronomy" in the
framework of which we have made a special laboratory facility able to
measure inerton waves. At present, we could record inerton signals
along the West-East line at $\sim 20$ Hz, which was associated with
proper rotation of the globe. From September to December, 2004, we
could record a flow of inertons at frequencies 18 to 22 kHz, which
came from the northern sky in a universal time interval from 3 p.m. to 5 p.m.

The concept of the tessel-lattice of space replaces such uncertain
notions as a classical elastic aether and a physical vacuum. This
deeper concept allows an uncovering of many inner details of the
constitution and behavior of particles and physical fields, which
still elude researchers.

\subsection*{References}

\small

\medskip \noindent Azjukowski, W. A., ed. (1993),\textit{ Aetheral
wind}, Collection of papers, Energoatomizdat, Moscow, 289 pages;
in Russian.

\bigskip\noindent Baurov, Yu. A. (2002), Structure of physical
space and nature of de Broglie wave (theory and experiment),
\textit{Annales de la Fondation Louis de Broglie} {\bf27}, pp.
443-461.

\bigskip\noindent Benford, M. S. (2002), Probable axion detection
via consistent radiographic findings after exposure to a Shpilman
axion generator, \textit{Journal of Theoretics} {\bf 4}, no. 1; \
http://www.journaloftheoretics.com/Articles/4-1/Benford-axion.htm

\bigskip \noindent Bounias, M. (2000), The theory of something: a
theorem supporting the conditions for existence of a physical
universe, from the empty set to the biological self,
\textit{International Journal of Computing Anticipatory Systems}
{\bf 5}, pp. 11-24.

\bigskip \noindent Bounias, M. and Krasnoholovets, V. (2003a),
Scanning the structure of ill-known spaces: Part 1. Founding
principles about mathematical constitution of space,
\textit{Kybernetes: The International Journal of Systems and
Cybernetics} {\bf 32}, no. 7/8, pp. 945-975 (also \
physics/0211096).

\bigskip \noindent Bounias, M. and Krasnoholovets, V. (2003b),
Scanning the structure of ill-known spaces: Part 2. Principles of
construction of physical space, \textit{Kybernetes: The
International Journal of Systems and Cybernetics} {\bf 32}, no.
7/8, pp. 976-1004 (also physics/0212004).

\bigskip \noindent Bounias, M. and Krasnoholovets, V. (2003c),
Scanning the structure of ill-known spaces: Part 3. Distribution
of topological structures at elementary and cosmic scales,
\textit{Kybernetes: The International Journal of Systems and
Cybernetics}  {\bf 32}, no. 7/8, pp. 1005-1020 (also
physics/0301049).

\bigskip \noindent Bounias, M. and Krasnoholovets, V. (2004), The
universe from nothing: A mathematical lattice of empty sets,
\textit{International Journal of Anticipatory Computing Systems},
ed. D. Dubois, in press (also physics/0309102).

\bigskip \noindent Bounias, M. and Krasnoholovets, V. (2002),
Science addendum to: Harezi, I. (2002), \textit{The resonance in
residence. An inner and outer quantum journey}, Ilonka Harezi,
USA.

\bigskip \noindent Bonaly, A. and Bounias, M. (1996), On metrics
and scaling: physical coordinates in topological spaces,\textit{
Indian Journal of Theoretical Physics} {\bf 44}, no. 4, pp.
303-321.

\bigskip
\noindent Einstein, A. (1905) Zur Elektrodynamik der bewegter
K\"orper, \textit{Annalen der Physik} {\bf 17}, SS. 891-921.

\bigskip \noindent Einstein, A. (1916), Die Grundlage der allgemeinen
Relativit\"atstheories, \textit{Annen der Physik} {\bf 49}, SS.
769-822.

\bigskip \noindent Einstein, A. (1920), {\"Aether} und
Relativit\"atstheorie, \textit{Lecture presented on
5}$^{th}$\textit{ May 1920 in the University of Leiden}, Springer,
Berlin.

\bigskip\noindent Essen, L. (1955), A new aether drift experiment,
\textit{Nature} {\bf 175}, pp. 793-794.

\bigskip \noindent Galaev, Yu. M. (2002), The measuring of
aether-drift velocity and kinematic aether viscosity within
optical waves band, \textit{Spacetime \& Substance} {\bf 3}, pp.
207-224.

\bigskip \noindent Kozyrev N. A. and Nasonov, V. V. (1978),
A new method of the determination of trigonometric parallaxes
based on the measurement of a difference between the true and
apparent star positions, in: \textit{Asrometry and Celestial
Mechanics}, Akademiya Nauk SSSR, Moscow, Leningrad, pp. 168-179;
in Russian.

\bigskip \noindent Krasnoholovets, V. (2002a), Submicroscopic
deterministic quantum mechanics, \textit{International Journal of
Computing Anticipatory Systems} {\bf 11}, pp. 164-179 (2002) (also
quant-ph/0109012).

\bigskip \noindent Krasnoholovets, V. (2002b), Gravitation as deduced
from submicroscopic quantum mechanics, submitted (also
hep-th/0205196).

\bigskip \noindent Krasnoholovets V., (2004), On the origin of conceptual
difficulties of quantum mechanics, in {\it Developments in Quantum
Physics}, eds. F. Columbus and V. Krasnoholovets, Nova Science
Publishers, New York, pp. 85-109.

\bigskip \noindent Miller, D. C. (1933), The aether-drift experiment and
the determination of the absolute motion of the Earth,
\textit{Review of Modern Physics} {\bf 5}, no. 3, pp. 203-242.

\bigskip \noindent Poincar$\acute{\rm e}$, H. (1906), Sur la dynamique de
l$\acute{\rm e}$lectron, \textit{Rendiconti del Circolo matematico
di Palermo} {\bf 21}, pp. 129-176; also: Oeuvres, t. IX, pp.
494-550 (also in Russian translation: Poincar$\acute{\rm e}$, H.
(1974), \textit{Selected Transactions}, ed. N. N. Bogolubov,
Nauka, Moscow {\bf 3}, pp. 429-486.

\bigskip  \noindent Poincar$\acute{\rm e}$, H. (1905), \textit{La valeur
de la science}, Flammarion, Paris; (1906), \textit{La science et
l'hypoth$\rm\grave{e}$èse} \textit{Science et m$\rm
\grave{e}$thode}, Flammarion, Paris; (1913), \textit{Derni$\rm
\grave{e}$res pens$\acute{\rm e}$es}, Flammarion, Paris. In
Russian translation: Poincar$\acute{\rm e}$, H. (1980),\textit{
About science}, Mir, Moscow).

\bigskip  \noindent Shipov, G. I. (1997), \textit{The theory of physical
vacuum}, Nauka, Moscow, pp. 247-296.

\bigskip
\noindent Urutskoev, L. I., Liksonov, V. I. and Tsioev, V. G.
(2002), Observation of transformation of chemical elements during
electric discharge, \textit{Annales de la Fondation Louis de
Broglie} {\bf 27}, pp. 701-726.

\end{document}